\documentclass[showpacs,preprintnumbers,amsmath,amssymb,12pt,prl,aps]{revtex4}
\usepackage{amsmath,amssymb}
\usepackage{epsfig}
\usepackage{citesort}
\usepackage{graphicx}
\usepackage{times}
\def\piby4{\frac{\pi}{4}}

\def\etal{\mbox{\it et al.}}

\newcommand{\bg}{\boldsymbol}

\newcommand{\be}{\begin{equation}}
\newcommand{\ee}{\end{equation}}

\begin{document}

\title{Orientational correlation and  velocity distributions in 
uniform shear flow of a dilute granular gas}
\author{Bishakdatta Gayen$^{1}$ and Meheboob Alam$^{1,2}$\footnote{Author to whom
correspondence should be addressed: meheboob@jncasr.ac.in;\\
To be published in {\it Physical Review Letters}, vol. 100, (2008)}
}
\affiliation{
$^{1}$Engineering Mechanics Unit, $^{2}$Max Planck Partner Group of MPI-Bremen, 
Jawaharlal Nehru Center for Advanced
Scientific Research, Jakkur PO, Bangalore 560064, India
}

\date{\today}

\begin{abstract}
Using particle simulations of the uniform shear 
flow of a rough dilute granular gas, we show that 
the translational and rotational velocities are strongly correlated in direction,
but there is no orientational correlation-induced singularity at perfectly 
smooth ($\beta=-1$) and rough ($\beta=1$) limits for elastic collisions ($e=1$);
both the translational and rotational velocity distribution functions 
remain close to a Gaussian for these two limiting cases.
Away from these two limits, the orientational as well as spatial velocity correlations
are  responsible for the emergence of non-Gaussian high velocity tails.
The tails of both distribution functions follow stretched exponentials,
with the exponents depending on  
normal ($e$) and tangential ($\beta$) restitution coefficients.
\end{abstract}

\pacs{45.70.Mg, 47.45.-n, 45.70.-n}

\maketitle

Under external forcing (e.g.  shearing/vibration),
the granular materials, a collection of macroscopic solid particles, 
can flow like a gas or a liquid.
The rapid flow of granular gases (dilute limit) has been extensively studied 
using kinetic theory~\cite{reviews1,Lun91,HZ97} that takes into
account the dissipative nature of particle interactions.
Most theories of granular gases assume that the particles are {\it smooth}
which is a simplification of real particles which
are always {\it rough}, giving rise to surface friction,
and the rotational motion is important to deal with such rough, frictional particles.
Prior literature~\cite{Lun91,HZ97,GNB05,BPKZ07,Simu-Dist1}
suggests that the rotational motion should not
be neglected for a realistic modeling of the dynamics and pattern
formation in a granular gas even in the Boltzmann (dilute) limit.
For example, the molecular dynamics (MD) simulations\cite{Simu-Dist1} 
have elucidated the crucial role of
friction on pattern formation in oscillated granular layers via a comparison
with experimental results.
Despite the importance of frictional interactions, only a small-body
of work exists on the modeling of rough granular gas~\cite{Lun91,HZ97,GNB05,BPKZ07}.

To develop constitutive models of rough granular gases,
a systematic study of correlations and the distribution functions
of both `translational' and `rotational' velocities is of fundamental interest.
While the deviation of translational velocity distribution functions (VDF)
from a Gaussian has been extensively studied 
(in terms of stretched exponential or power-law tails)
using theory~\cite{Theo-Dist}, simulation~\cite{Simu-Dist} and experiment~\cite{Expt-Dist},
similar results on `rotational' VDFs are very scarce.

For a `rough' granular gas, one needs to probe possible `orientational'/`directional' 
correlations  between translation and rotation,
in addition to standard density and velocity correlations.
It has been recently found~\cite{BPKZ07} that such orientational correlations are strong and 
the limit of smooth granular gas is {\it singular} in a freely cooling granular gas.
The last result readily raises doubts about the validity of the
hydrodynamic theories~\cite{GNB05} that are obtained 
via  perturbative expansions around the smooth-particle limit.
Also, it is of interest to ascertain  the impact of  such orientational correlation
on VDFs, especially in two limits of perfectly smooth and rough
particles around which the perturbative expansions are sought.
The above issues are investigated in this paper
using MD simulations of the {\it uniform shear} flow (which is  
a prototype non-equilibrium steady-state to 
develop constitutive relations\cite{GNB05}) of a dilute rough granular gas.

We consider a mono-disperse system of {\it rough}, inelastic
spheres of diameter $d$, mass $m$, and the moment of inertia ${\mathcal I}$,
interacting via purely repulsive potential.
Let us denote the {\it pre-collisional} translational and rotational velocities
of particle $i$  by ${\bf c}_i$ and ${\bg\omega}_i$, respectively,
and the corresponding {\it post-collisional}  velocities are denoted by the primed symbols,
${\bf c}'_i$ and ${\bg\omega}'_i$.  The pre-collisional relative velocity
at contact, $\bf g_{ij}$, between particle $i$ and $j$ is 
${\bf g}_{ij} = {\bf c}_{ij} - (d/2){\bf k}{\bg\times} ({\bg\omega}_i + {\bg\omega}_j)$,
where ${\bf c}_{ij} = {\bf c}_i - {\bf c}_j$ is the relative translational velocity  
between particle $i$ and  $j$, and ${\bf k}$ is the unit vector directed from the 
center of particle $j$ to that of particle $i$.
Neglecting Coulomb friction in the dilute limit,
we have the following collision model\cite{Lun91}:
\[
{\bf k}{\bg\cdot}{\bf g}'_{ij} = - e({\bf k}{\bg\cdot}{\bf g}_{ij}) 
\quad \mbox{and} \quad 
 {\bf k}{\bg\times}{\bf g}'_{ij} = - \beta({\bf k}{\bg\times}{\bf g}_{ij}), 
\]
characterized by two parameters:
the normal restitution coefficient, $0\leq e \leq 1$, 
and the tangential restitution coefficient, $-1 \leq \beta \leq 1$. 
The former is an indicator of the {\it inelasticity} of
particle-collisions and the latter characterizes its surface roughness.
For collisions between  perfectly smooth particles $\beta=-1$,
with increasing value of $\beta$ being an indicator of the
increasing degrees of particle surface friction.
The value of $\beta=0$ represent the case for which
the particle surface friction and inelasticity are sufficient to
eliminate the post-collisional tangential relative velocities.
For $0< \beta \leq 1$, the {\it spin-reversal} occurs after 
collision\cite{Lun91}, and the case of $\beta=1$ 
corresponds to collisions between perfectly rough particles.

We have used an event-driven algorithm~\cite{Lubachev91} 
to simulate the uniform shear flow, characterized by a linear velocity profile.
The simulation box is a cube, with Lees-Edwards boundary condition~\cite{Lees72}
across two moving boundaries along $y$-direction, 
and the periodic boundary conditions along $x$- and $z$-directions,
with $x$ being the direction of flow.
The positions of particles are initialized in the simulation box,
with their translational and rotational velocities being 
chosen randomly from a Gaussian distribution.
The data are accumulated once the system has reached a  statistical steady-state condition
which is monitored from the temporal evolution of the system's kinetic energy. 
The coarse-grained 'translational' fluctuation kinetic energy (i.e.
the standard granular temperature ), $T$, and the `rotational' fluctuation 
kinetic energy, $\theta$, are defined as: 
$T({\bf x}, t) = \langle {\bf C}\cdot {\bf C}\rangle/3 $ 
and
$\theta({\bf x}, t) = ({\mathcal I}/{3m})\langle \Omega\cdot \Omega\rangle$,
respectively, where the angular bracket denotes a suitable averaging.
Here ${\bf C}= {\bf c} - \langle{\bf c}\rangle$ is the translational `peculiar' velocity  
which measures the deviation of the instantaneous particle velocity 
from the local mean velocity ($\langle{\bf c}\rangle$),
and $\Omega = {\bg\omega} - \langle{\bg\omega}\rangle$ is its rotational counterpart.
All results are presented for a very dilute system (Boltzmann limit),
with a volume fraction of $\phi=0.01$.
(The effect of density, as  well as Coulomb friction, is a
non-trivial issue which will be taken up  in a separate publication.)
The number of particles was fixed to $N=8000$, and the robustness of results was
checked by using $N=4000$ and $16000$.

\begin{figure}[h!]
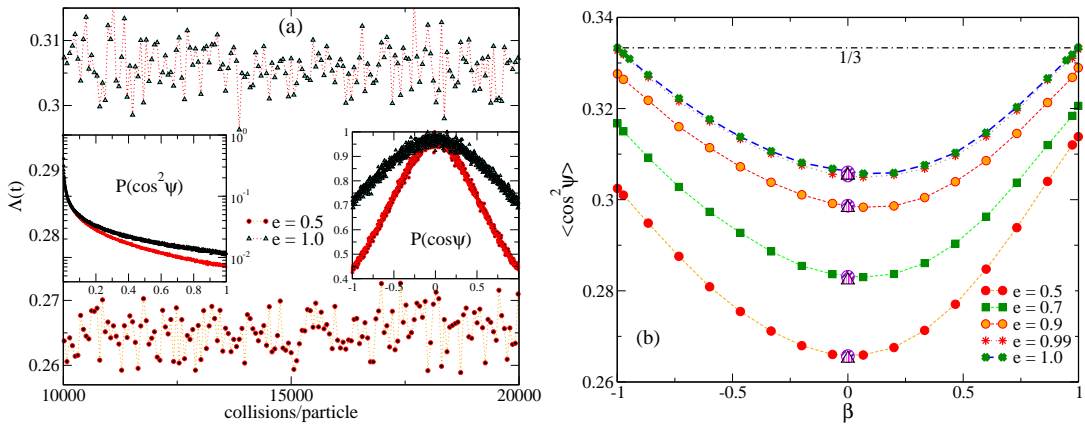

\includegraphics[width=2.8in]{figure1a.eps}
\includegraphics[width=2.8in]{figure1b.eps}
\caption{(color online)
($a$) Variation of $\Lambda(t)=\cos^2\Psi$ with time for different 
values of $e$ with $\beta=0$, $\phi = 0.01$ and  $N=8000$. 
Left and right insets show the distributions of $\cos^2\Psi$
and $\cos\Psi$, respectively.
($b$) Variation of $\langle\cos^2\Psi\rangle$ with  $\beta$ for different $e$.
Larger symbols (triangle and hatched-circle) at $\beta=0$ for each $e$ correspond to
simulations with $N=4000$ and $16000$, respectively.
}
\label{fig:fig1}
\end{figure}

It is known that the translational and rotational fluctuating velocities 
are uncorrelated in a molecular gas, but they have been shown to be correlated 
(in direction) in a freely cooling granular gas~\cite{BPKZ07}.
This orientational/directional
correlation between translational and rotational velocities is quantified in terms of the 
mean square of the cosine of the angle, $\Psi$, between 
$C={\bf c}-\langle{\bf c}\rangle$ and $\Omega={\bg\omega}-\langle{\bg\omega}\rangle$:
\begin{equation}
 \Lambda(t) = \frac{1}{N}\sum_{i=1}^{N}\frac{\left(C_i\cdot \Omega_i\right)^2}
           {\left(C_i^2 \Omega_i^2\right)}
       = \frac{1}{N}\sum_{i=1}^{N}\cos^2\Psi_i \equiv \cos^2\Psi .
\end{equation}
In fig.~\ref{fig:fig1}($a$) we have plotted the temporal variation
of $\Lambda(t)$ (main panel) for two values of normal restitution coefficients ($e=1, 0.5$),
with the tangential restitution coefficient being set to $\beta=0$;
the corresponding probability distribution of $\cos^2\Psi$ is shown in the left inset.
(The probability distribution of $\cos\Psi$, $P(\cos\Psi)$, is 
symmetric about its zero mean for all $e$, but its width becomes narrower
with decreasing $e$, see the right inset.)
From the main panel and the left inset, we find  that even for $e=1$ the
mean value of $\Lambda$ is different from $1/3$ (for a molecular gas),
signaling the presence of {\it orientational/directional} correlation;
decreasing the value of $e$ to $0.5$ decreases its value to $\langle\Lambda\rangle\sim 0.26$,
thus enhancing orientational correlation significantly.
The variation of the temporal-average of 
$\Lambda(t)$ with particle roughness, $\beta$,
is shown in fig.~\ref{fig:fig1}($b$); the dot-dash line represents the limiting
value of $1/3$ for a molecular gas.
Note that the data points for $e=1$ (thick blue dashed line) 
and $e=0.99$ almost overlap with each other. 
For any $e$, the orientational correlation is maximum at $\beta\sim 0$ and
it decreases {\it monotonically} as we approach the perfectly smooth ($\beta=-1$)
and perfectly rough ($\beta=1$) limits. This latter observation is in contrast
to that in a freely cooling dilute granular gas~\cite{BPKZ07} for which
$\langle\Lambda\rangle$ varies non-monotonically with $\beta$ for $-1<\beta<0$ and $0<\beta<1$.
Another difference with freely cooling gas is that the magnitude of
$\langle\Lambda\rangle$ is much larger in shear flow.
It must be noted that even though the translation and rotation are decoupled at $\beta=-1$
(independent of the value  of $e$),
the smooth limit is singular for any $e\neq 1$ in shear flow.
However, there is no orientational correlation-induced singularity at
both the perfectly smooth ($\beta=-1$) and rough ($\beta=1$) limits
for the limiting case of $e=1$.

\begin{figure}[!ht]
\centering
\includegraphics[width=2.8in]{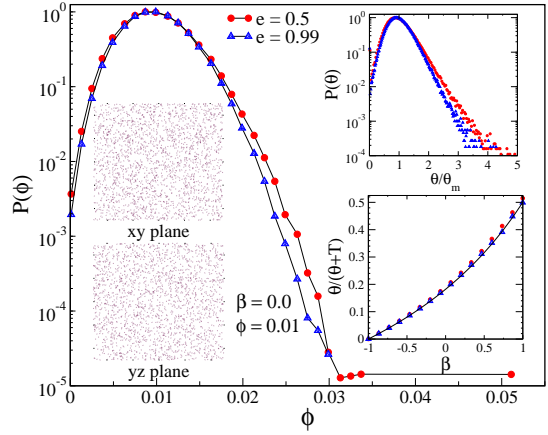}
\caption{(color online)
Probability distributions of mean density (main panel) and rotational temperature 
(right upper inset) for different values of $e$.
Right lower  inset shows the variation of temperature ratio, $\theta/(\theta+T)$, with $\beta$.
Two left insets show projected particle snapshots in the $xy$- and $yz$-planes
at steady states for $e=0.5$.
}
\label{fig:fig2}
\end{figure}

Before presenting results on VDFs,
we  probe mean-field quantities to ascertain the presence of any
inhomogeneity in our system.
Figure \ref{fig:fig2} shows the probability distributions of
the mean density (main panel) and the rotational temperature
(right upper inset) for  $\phi=0.01$ and  $\beta=0$;
the distribution of mean translational temperature ($T$) looks similar
to that of $\theta$ (not shown).
The lower right inset shows that the translational and rotational temperatures
are un-equally partitioned over the whole range of $\beta$ 
(except at $\beta=1$ with $e=1$), and the calculated temperature ratio (symbols)
agrees well with theoretical predictions~\cite{Lun91,HZ97} (solid line).
The data for mean distributions in Fig.~2 have been  obtained by dividing the simulation box 
into a number of equal-sized cells ($10^3$) such that on average  
about ten particles occupy  each cell and then calculating the 
instantaneous value of any mean field quantity 
($\phi$, $T$, $\theta$, $\langle{\bf c}\rangle$) in each cell.
It is interesting that even though the mean density varies 
between $0.003$ and $0.017$ in about $90\%$ cells,
the density-distribution remains almost identical with decreasing $e$ from $0.99$ to $0.5$.
The projected snapshots of all particles (at steady state after 
$60000$ collisions per particle) in the $xy$- and $yz$-planes,
as displayed in two left insets of Fig.~2, further suggest that
the particles are  homogeneously distributed and 
there is no discernible dissipation-induced clustering even at $e=0.5$ in our system.

\begin{figure}[h!]
\includegraphics[width=2.8in,height=2.8in]{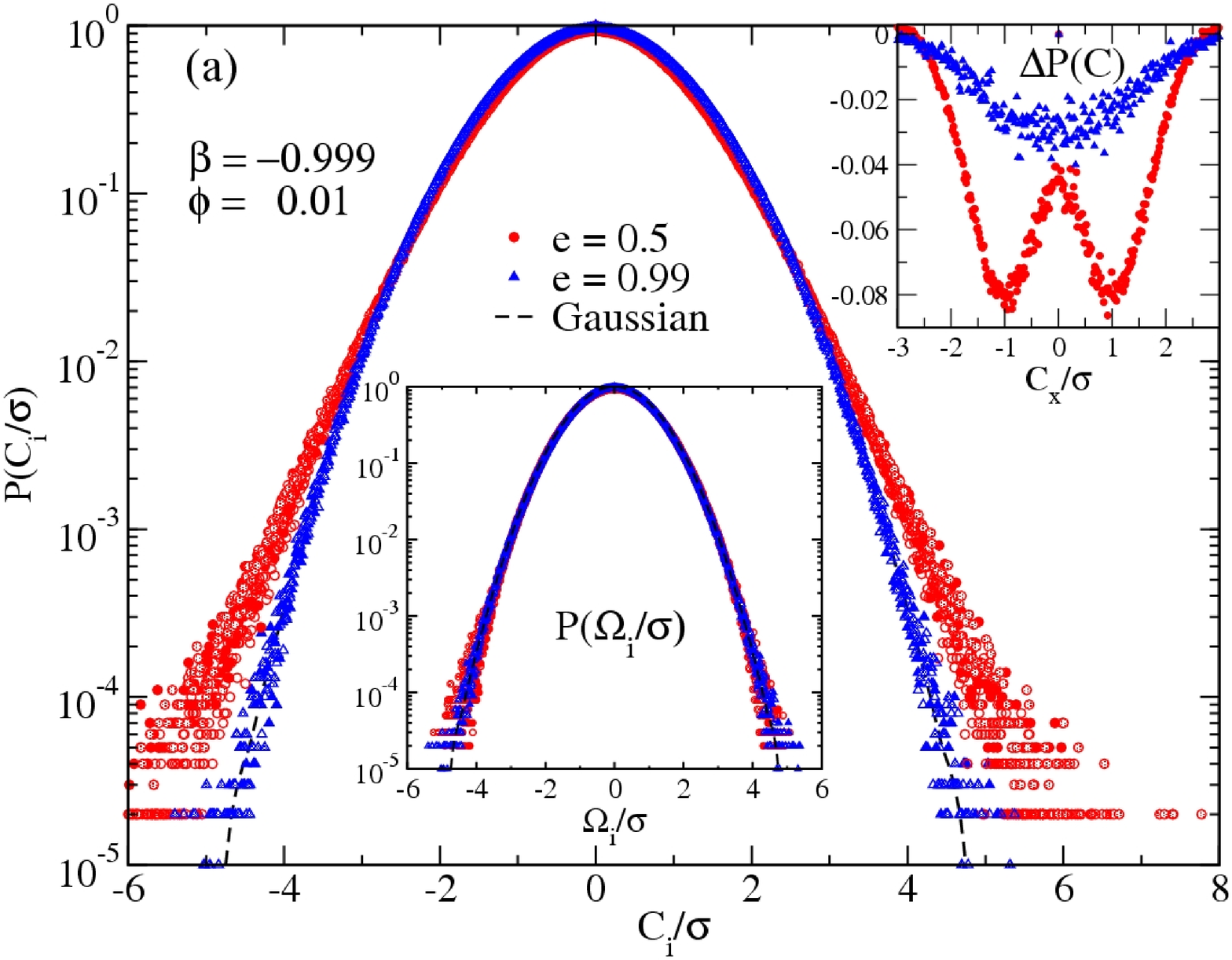}
\includegraphics[width=2.8in,height=2.8in]{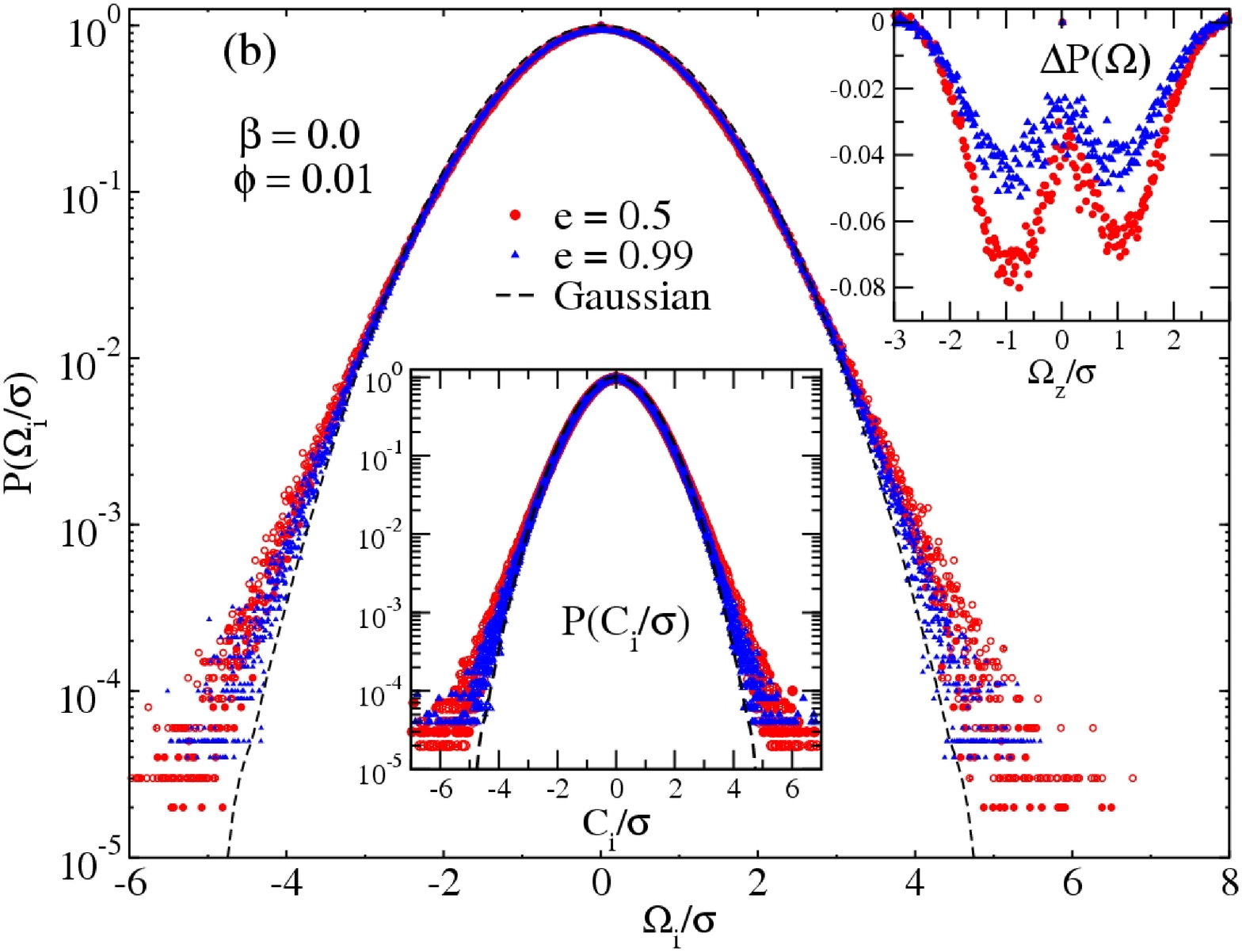}
\caption{(color online)
($a$) Translational (main panel) and rotational (lower inset) VDFs for $\beta=-0.999$.
($b$) Rotational  (main panel) and translational (lower inset) VDFs for $\beta=0$.
The upper right inset in each panel shows the deviation of the low
velocity regions from a Gaussian.
}
\label{fig:fig3}
\end{figure}

Now  we turn to  velocity distribution functions which are
known to be strongly affected by the presence of 
correlations~\cite{Simu-Dist,Theo-Dist,Expt-Dist}.
To calculate VDFs, we used  {\it cell-wise} averaging as discussed above.
Figure~3($a$) shows the probability distribution functions of the translational (main panel) 
and rotational (lower inset) fluctuating velocities for $\beta=-0.999$,
while Fig.~3($b$) shows the translational (lower inset) and rotational (main panel)
VDFs for $\beta=0$.  In each plot, two data sets for $e=0.99$ and $0.5$ 
have been superimposed, and the black dashed line represents a Gaussian.
The horizontal axis of each plot has been scaled by $\sigma$,
the standard deviation of the corresponding distribution,
and the vertical axis scaled such that $P(0)=1$. 
For the perfectly smooth limit ($\beta\sim -1$) in Fig.~3($a$),
the tails of the translational VDFs deviate
from a Gaussian with increasing dissipation; the rotational
VDFs remain a Gaussian  even for $e=0.5$.
We did not find any discernible difference among the VDFs
for all three components of each velocity ($C_i$ and $\Omega_i$, with $i=x,y,z$)
and that they follow the same distribution for a given $e$.
For the other extreme of perfectly rough limit ($\beta\sim 1$, not shown),
the translational VDFs follow a similar behavior as that for $\beta\sim -1$, 
but the rotational VDFs become non-Gaussian with increasing dissipation.
At intermediate values of roughness ($\beta=0$) as in Fig.~3($b$),
both the translational and rotational velocity distributions deviate
from a Gaussian even in the quasi-elastic limit ($e\sim 1$).
It is seen that both $P(C_x)$ and $P(\Omega_z)$ are under-populated for low-velocities
and over-populated for high-velocity tails.
The upper right insets in Figs.~3($a$) and 3($b$) show the deviation of low
velocities from a Gaussian
(i.e. $\triangle P(x) = P(x)-\exp(-x^2/2)$, with $x=C/\sigma, \Omega/\sigma$). 
It is noteworthy that the low-velocity region of $P(\Omega_z)$ is
slightly asymmetric for $e=0.5$; such asymmetry is absent for $\Omega_x$
and $\Omega_y$ and at higher values of $e>0.7$.
Such asymmetry implies the onset of preferential transport of rotational velocity fluctuation along
the negative $z$-direction (i.e. the mean vorticity direction of the steady shear flow), 
but the reason for the emergence of this asymmetry 
(at large dissipation levels) remains unclear with periodic boundary conditions along $z$.

\begin{figure}[h!]
\includegraphics[width=2.8in,height=2.8in]{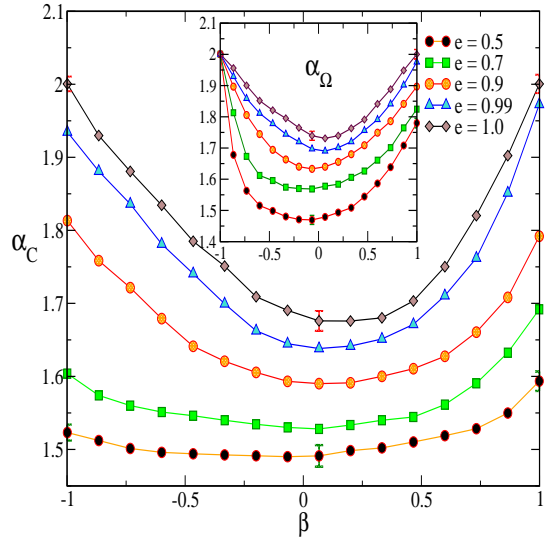}
\caption{(color online)
Variation of the exponent for the stretched exponential, $\alpha_i$, with $\beta$;
translational (main panel, $\alpha_C$) and rotational (inset, $\alpha_\Omega$) velocity.
Typical error-bars are shown on few data points. 
}
\label{fig:fig4}
\end{figure}

Our data on high-velocity tails  of $P(C)$ and $P(\Omega)$ 
have been  fitted with  stretched exponential functions~\cite{Theo-Dist} of the 
form $P(x) \sim \exp(-\gamma_x x^{\alpha_x})$, 
where $\alpha_x$ and $\gamma_x$ (with $x=C, \Omega$)
are the exponent and pre-factor of the corresponding distribution. 
Figure~4 shows the variations of $\alpha_C$ (main panel)
and $\alpha_\Omega$ (inset) with roughness $\beta$ for different $e$. 
For $e=1$ the tails of both translational and rotational VDFs deviate from a Gaussian
(except at  $\beta=\pm 1$ for which $\alpha_C=2=\alpha_\Omega$),
and  the functional-forms of $\alpha_C$
and $\alpha_\Omega$ are asymmetric and symmetric (around $\beta=0$), respectively.
With increasing inelasticity, both $\alpha_C$ and $\alpha_\Omega$ decrease sharply,
and $\alpha_\Omega$ also becomes asymmetric around $\beta=0$.
A least-square fit to our data suggests that $\alpha_i$ follows a power-law relation
with inelasticity: $\alpha_i = 2 - A_i(1-e^2)^{B_i}$ with $i=C, \Omega$,
and  $(A_C, B_C)\approx (5/8, 2/3)$ at $\beta=- 1$, 
and $(A_C, B_C)\approx (3/4, 1)$ and $(A_\Omega, B_\Omega) \approx (3/8, 7/8)$ at $\beta=1$.
Due to the asymmetry of $\alpha_C$ and $\alpha_\Omega$ around $\beta=0$,
we could not find an universal scaling of $\alpha_C$ and $\alpha_\Omega$ with
$(1-\beta^2)$ at any $e$.

\begin{figure}[!ht]
\includegraphics[width=2.8in,height=2.8in]{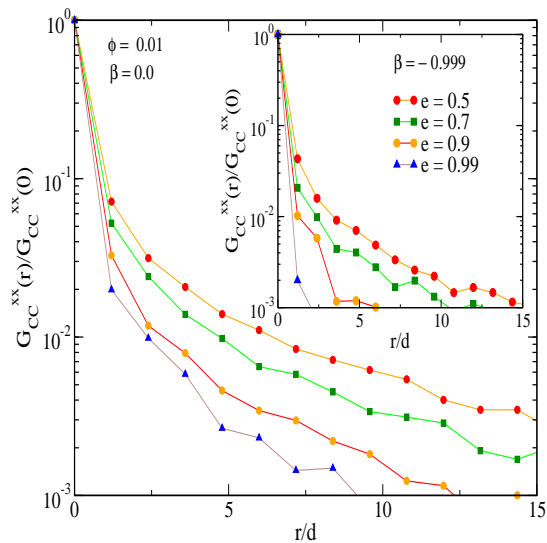}
\caption{(color online)
Spatial velocity correlation: $\beta = 0$ (main panel)
and $\beta=-0.999$ (inset).
}
\label{fig:fig5}
\end{figure}

Lastly, the variations of spatial translational velocity correlation function,
$G_{CC}^{xx}(r/d)= \langle C_x(R)C_x(R+r)\rangle$, 
for different $e$ are shown in fig.~5 for  $\beta=0$ (main panel)
and $\beta=-0.999$ (inset)-- other velocity components follow a similar behavior.
It is seen that   $G_{CC}^{xx}$ (and related correlation length)
increases  with increasing inelasticity for a given $\beta$,
and the effect of  rotational dissipation ($\beta$)
is more prominent in the quasi-elastic limit ($e\sim 1$).
(We have checked that the pair correlation function, $g(r/d)$, for any $\beta$ (not shown)
is featureless, except that its contact value increases slightly with decreasing $e$.)
A general finding is that in the quasi-elastic limit the density and spatial
velocity correlations are negligible near $\beta=\pm 1$,
however, the spatial correlations for translational velocities emerge with
increasing rotational dissipation and becomes strong near $\beta=0$.
These velocity correlations, together with orientational correlations, are responsible for
non-Gaussian VDFs in a rough granular gas.

In conclusion, we showed that the translational and rotational velocities 
are directionally correlated in a dilute sheared granular gas, 
but there is no orientational correlation-induced singularity 
at both the perfectly smooth ($\beta=-1$) and rough ($\beta=1$) limits for $e=1$. 
Even though the translational and rotational degrees of freedom are
uncorrelated at $\beta=-1$ for any value of $e$, the smooth-limit ($\beta\to -1$)
is singular (i.e. $<\Lambda>\neq 0$) for any $e\neq 1$.
Both `translational' and `rotational' VDFs remain close to a Gaussian 
for these two limiting cases; away from $\beta=\pm 1$, the orientational 
correlations and spatial velocity correlations
are responsible for the emergence of non-Gaussian VDFs.
The tails of both VDFs follow stretched exponentials,
with the exponents depending on two restitution coefficients ($e$ and $\beta$).
One immediate consequence of our results is the resolution
of a doubt~\cite{BPKZ07} that in a dilute sheared granular gas
the perturbative expansions around the smooth (and also rough) limit
is appropriate with Gaussian being the leading-order VDF~\cite{GNB05}.

{\bf Acknowledgement:} We acknowlege partial funding support from the Max-Planck Society, Germany,
for the `Max-Planck Partner Group for Topography Formation' at JNCASR.


\end{document}